# Applicability and Limitation Analysis of PMU Data and Phasor Concept for Low- and High-Frequency Oscillations

Bowen Ou, *Graduate Student Member, IEEE*, Bin Wang, *Senior Member, IEEE*, Slava Maslennikov, *Fellow, IEEE*, Hanchao Liu, *Senior Member, IEEE*, Jim Follum, *Senior Member, IEEE*

*Abstract*—**Phasor Measurement Units (PMUs) convert high-speed waveform data into low-speed phasor data, which are fundamental to wide-area monitoring and control in power systems, with oscillation detection and localization among their most prominent applications. However, representing electrical waveform signals with oscillations using PMU phasors is effective only for low-frequency oscillations. This paper investigates the root causes of this limitation, focusing on errors introduced by Discrete Fourier Transform (DFT)-based signal processing, in addition to the attenuation effects of anti-aliasing filters, and the impact of low reporting rates. To better represent and estimate waveform signals with oscillations, we propose a more general signal model and a multi-step estimation method that leverages one-cycle DFT, the Matrix Pencil Method, and the Least Squares Method. Numerical experiments demonstrate the superior performance of the proposed signal model and estimation method. Furthermore, this paper reveals that the phasor concept, let alone PMU phasors, can become invalid for waveform signals with high-frequency oscillations characterized by asymmetric sub- and super-synchronous components. These findings highlight the fundamental limitations of PMU data and phasor concept, and emphasize the need to rely on waveform data for analyzing high-frequency oscillations in modern power systems.**

## I. INTRODUCTION

P HASOR Measurement Units (PMUs) are widely deployed in power systems to enhance situational awareness for both real-time and offline applications [1], [2]. In the context of power system oscillation monitoring [3], [4], PMUs have achieved notable success [5], [6]. For instance, reference [7] evaluated the relative performance of P-Class and M-Class PMUs by examining their responses to low-frequency inter-harmonic signals and analyzing total vector error under modulation frequencies below 5 Hz. However, the performance of PMU algorithms in estimating high-frequency oscillations has received limited attention until the recent study in report [8] which briefly states that PMU data lacks sufficient accuracy for capturing high-frequency oscillations. The inaccuracies are primarily attributed to signal attenuation caused by filters at frequencies above 5 Hz and to aliasing effects resulting from low reporting rates. However, report [8] does not provide any details about such a limitation or other possible limitations.

Sub-synchronous oscillation events associated with inverter-based resources (IBRs) have been reported over a broad frequency spectrum, ranging from approximately 0.1 Hz to several kilohertz [9]- [10], often exhibiting fast and complex dynamic behavior. As the oscillation frequency increases, the accuracy of PMU-based monitoring may be compromised. This has motivated research in analyzing PMU limitations. For example, reference [11], using hardware in the loop experiments, revealed several notable effects, including aliasing in the single sided frequency spectrum by comparing PMU outputs with root mean square signals. The results suggest that filtering and signal processing stages may significantly degrade PMU measurement accuracy of higher-frequency components. Similarly, reference [12] reported that PMU measurements remain reliable only for oscillation frequencies up to half the reporting rate and are susceptible to aliasing when this limit is exceeded.

Still, it is not clear how significant the inaccuracies are when observing high-frequency oscillations using PMU data, and if inaccuracies can become significant, how to avoid them. To fill in this gap, this paper presents a comprehensive investigation, with main contributions summarized as follows:

- Based on a typical PMU algorithm, we identify the three main error sources of PMU data, i.e., DFT algorithm, anti-aliasing filter, and the reporting rate.
- We show analytically that the DFT algorithm introduces errors to the estimated oscillation amplitude and oscillation phase. Specifically, we show that the DFT-induced error depends on the length of time window and the underlying oscillation frequency. The longer the time window or the higher the

This material is based upon work supported by the U.S. Department of Energy's Office of Energy Efficiency and Renewable Energy (EERE) under the Solar Energy Technologies Office Award Number 52372. The views expressed herein do not necessarily represent the views of the U.S. Department of Energy or the United States Government.

Paper no. TBD. (*Corresponding author: Bin Wang.*)

Bowen Ou is with the University of Texas at San Antonio, San Antonio, TX 78249 USA (email: bowen.ou@my.utsa.edu).

Bin Wang is with the University of Texas at San Antonio, San Antonio, TX 78249 USA, and also with ISO New England, Holyoke, MA 01040 USA (email: bwang@iso-ne.com).

Slava Maslennikov is with ISO New England, Holyoke, MA 01040 USA (email: smaslennikov@iso-ne.com).

Hanchao Liu is with GE Vernova, Schenectady, NY 12345 USA (email: hanchao.liu@ge.com).

Jim Follum is with the Pacific Northwest National Laboratory, Richland, WA 99354 USA (email: james.follum@pnnl.gov).

oscillation frequency, the greater the error will be.

- We propose a more general signal model to better represent waveform signals with high-frequency oscillations and a multi-step estimation method to estimate all parameters in the signal model.
- We reveal that both PMU phasor and phasor concept itself can become invalid for waveform signals with high-frequency oscillations characterized by asymmetric sub- and super-synchronous components, which are common for power system oscillations driven by inverter-based resources.

The remainder of the paper is organized as follows. Section II uses numerical examples to show the limitations of typical DFT-based phasor measurements. Section III presents a theoretical investigation into the limitations and applicability of DFT-based phasors. Section IV proposes a multi-step estimation method to accurately represent the point-on-wave data with either low- or high-frequency oscillations. Section V summarizes the applicability of the phasor concept. Section VI presents case studies to validate the paper's findings, and Section VII gives conclusions.

## II. PROBLEMS OF DFT-BASED PMU PHASOR

This section first introduces a typical DFT-based PMU algorithm and then presents two examples demonstrating phasor inaccuracies for waveforms with high-frequency oscillations, where errors introduced by each of the three main steps in a typical PMU algorithm are illustrated.

Without loss of generality, only single-phase waveform signals are considered in this paper.

### A. Review of typical DFT-based PMU algorithm

Reference [13] outlines the fundamental phasor estimation algorithm. As shown in Fig. 1, the analog waveform signal is first sampled and digitized by an analog-to-digital converter (ADC), followed by a low-pass filter to suppress the noise. The resulting discrete point-on-wave (PoW) data is then processed by an $N$-cycle DFT to estimate the phasor. The obtained phasor is fed to the anti-aliasing filter to suppress aliasing components [7][14]. Then, the output of the filter is timestamped at the center of the $N$-cycle window [14]. Finally, the PMU phasor is reported at a predefined reporting rate [14][15].

It is well-known that the second low-pass filter, i.e. LPF2 in Fig. 1, introduces amplitude attenuation, while aliasing effects arise due to the low reporting rate [8]. However, the accuracy of DFT-based phasor estimation remains unclear, particularly in the presence of high-frequency oscillations. The next subsection will show the errors introduced by each of the above three steps when dealing with high-frequency oscillations.

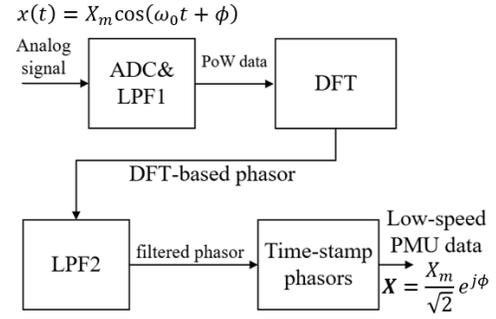

Fig. 1. Typical PMU algorithm based on DFT

### B. Illustration of PMU errors for high-frequency oscillations

Consider a voltage waveform with oscillations modulated in its amplitude:

$$V(t) = V_m(t) \cdot \cos(2\pi f_1 t + \varphi_1) \tag{1}$$

$$V_m(t) = A[1 + m\cos(2\pi f_{os} t + \varphi_2)] \tag{2}$$

where $t$ represents the time and $f_1$, $A$, and $\varphi_1$ are the fundamental frequency, amplitude and initial phase of the fundamental frequency component, respectively. The terms $f_{os}$, $m$, and $\varphi_2$ are the frequency, amplitude and initial phase of the modulated oscillations. To facilitate the comparison between phasor and waveform data, this paper uses amplitude rather than root mean square (RMS) magnitude for phasors. Note that amplitude is $\sqrt{2}$ times the RMS magnitude, whereas actual PMUs define phasors in terms of magnitude. In this paper, $f_1$=60 Hz in all numerical studies, while all derived formulas are also applicable for off-nominal frequencies.

Fig. 2 illustrates the PMU measurement process in detail using two waveform signals with $f_{os}$ = 2 Hz and $f_{os}$ = 40 Hz. Other parameters in both signals are identical: $\varphi_1$ = $\pi/4$, $A$ = 2, $m$ = 5%, $\varphi_2$ = $\pi/5$. Red curves represent the synthesized waveform signal $V(t)$ in (1) sampled at 2 kHz, blue curves represent the modulated amplitudes $V_m(t)$ in (2), which serve as the ground truth of phasor amplitude. Pink curves represent the phasor amplitude measured by the one-cycle DFT algorithm, green curves represent the measured phasor amplitude after LPF2 whose cut-off frequency is 60 Hz [13], and black curves with circle markers represent the measured phasor amplitude as PMU output with a typical reporting rate of 30 frames per second (fps).

In Fig. 2a, all measured phasor amplitude curves are nearly on top of each other, showing that the typical PMU algorithm can accurately capture the 2-Hz oscillation. However, when it comes to 40-Hz oscillation, each of the three steps in Section II-A introduce errors to the measurement. Specifically, DFT and LPF2 respectively attenuate the oscillation amplitude, while the low reporting rate causes aliasing, i.e., the 40-Hz oscillation appears as a 10-Hz oscillation when sampled at rate of 30 fps.

Comparing the two cases in Fig. 2, it is obvious that the typical DFT-based PMU algorithm can accurately characterize electromechanical oscillations usually with a frequency below 3 Hz, while significant errors may be introduced by the three main steps in PMU when dealing with sub-synchronous oscillations at 40 Hz: DFT, LPF2, and reporting. The impacts of LPF2 and reporting are explained in [8], which can be mitigated by increasing LPF2's cut-off frequency and increasing the reporting rate, though these adjustments involve tradeoffs in the PMU's overall performance [13]. However, the impact from DFT is not straightforward, which will be examined analytically and numerically in the next Section.

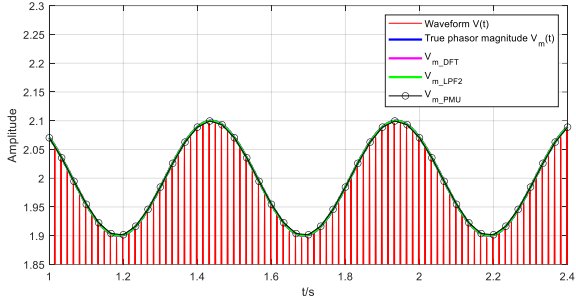

**(a)** 2-Hz oscillation case

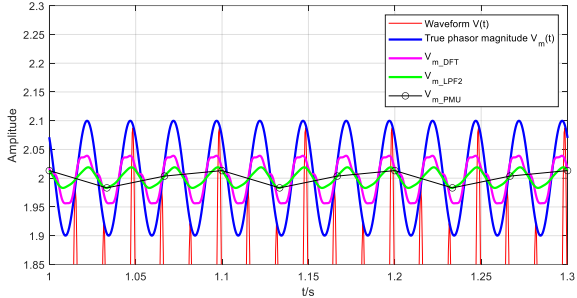

**(b)** 40-Hz oscillation case

Fig. 2. Measured phasor amplitude by a typical PMU algorithm

## III. APPLICABILITY AND LIMITATION OF DFT-BASED PHASOR

This section derives the analytic expression of DFT-based phasor for waveform signals containing amplitude-modulated oscillations. The explicit formulation provides insights into the estimation accuracy under both low- and high-frequency oscillations using different window lengths.

### A. DFT phasor for amplitude-modulated waveform signals

Consider the waveform signal model in (1)-(2). Applying an $N$-cycle Fourier transformation to time window $[\tau, \tau+N/f_1]$ gives

$$X(t) = \frac{2f_1}{N} \int_{\tau}^{\tau+\frac{N}{f_1}} V(t) e^{-j2\pi f_1 t} \, dt \triangleq X_r(t) + jX_i(t) \quad (3)$$

where $X_r(t)$ and $X_i(t)$ represent the real and imaginary parts of the DFT-based phasor expression, respectively. Note that the time stamp of the resulting phasor in (3) is chosen as the center of the window according to [14], i.e., $t = \tau+N/2f_1$. The reason will be explained in Section III-F.

Substituting (1)-(2) into (3), working out the integral, and obtaining $X(t)$'s amplitude as $X_m(t) = \sqrt{X_r^2(t) + X_i^2(t)}$ leads to

$$X_m(t) \approx A\left[1 + mF_{gain}\cos\left(2\pi f_{os}\tau + \frac{\pi f_{os}N}{f_1} + \varphi_2\right)\right] \quad (4)$$

$$F_{gain}(f_1, f_{os}, N) = \frac{2}{N} \frac{\sin\left(\frac{\pi f_{os}N}{f_1}\right)}{f_{os}\pi(4f_1^2 - f_{os}^2)}(2f_1^3 - f_{os}^2 f_1) \quad (5)$$

where $F_{gain}$ is named the measurement gain in this paper to quantify the distortion of DFT-measured oscillation amplitude, since the true oscillation amplitude $m$ in (2) is measured by DFT as $mF_{gain}$ in (4)-(5). Key derivation steps for (4)-(5) are included in Appendix A.

Eq. (5) shows that $F_{gain}$ is a function of $f_1$, $f_{os}$ and $N$. A plot of $F_{gain}$ as a function of these variables can be found in Section III. E. Usually, the synchronous frequency $f_1$ stays very close to its nominal value, and most PMUs mitigate the impact of off-nominal synchronous frequency by adopting a resampling technique [16] [17]. Thus, the two primary factors influencing the distortion of the DFT-based phasor are: (i) the oscillation frequency $f_{os}$, and (ii) the DFT window length $N$.

### B. DFT phasor for low frequency oscillations

For low frequency oscillations we have $F_{gain} \approx 1$ as shown in (6). The electromechanical oscillations below 3 Hz, the oscillation amplitude measured by the typical DFT-based PMU algorithm is fairly accurate, as illustrated in Fig. 2a.

$$\begin{aligned}
&\lim_{f_{os}\to 0} F_{gain}(f_1, f_{os}, N) \\
&= \lim_{f_{os}\to 0} \frac{2}{N} \frac{\sin\left(\frac{\pi f_{os}N}{f_1}\right)}{f_{os}\pi(4f_1^2 - f_{os}^2)}(2f_1^3 - f_{os}^2 f_1) \\
&= \frac{2}{\pi N} \cdot \frac{2f_1^3}{4f_1^2} \lim_{f_{os}\to 0} \frac{\sin\left(\frac{\pi f_{os}N}{f_1}\right)}{f_{os}} \\
&= \frac{2}{\pi N} \cdot \frac{f_1}{2} \cdot \frac{\pi N}{f_1} = 1
\end{aligned} \quad (6)$$

### C. DFT phasor for high frequency oscillations

For high-frequency oscillations whose frequencies are below $\sqrt{2}f_1$, we have (i) $4f_1^2 - f_{os}^2 > 4f_1^2 - 2f_1^2 = 2f_1^2 > 0$, (ii) $2f_1^3 - f_{os}^2 f_1 > 2f_1^3 - 2f_1^2 f_1 = 0$, and (iii) the sign of $F_{gain}$ in (5) depends only on the term $\sin(\pi f_{os}N/f_1)$. Specifically, when $\pi f_{os}N/f_1 = \pi$, i.e., $f_{os}=f_1/N$, this sine term will become zero, which means no matter how large the true oscillation amplitude $m$ is, the DFT-measured oscillation amplitude $mF_{gain}$ is always zero, i.e., the oscillation is completely invisible by PMU data. If $f_{os}$ is slightly greater than $f_1/N$, $F_{gain}$ will have a negative value and $-F_{gain}$ will be a positive number. In this case, according to the trigonometric identity in (7), the DFT-measured oscillation is out of phase with the underlying true oscillation. This analytically predicted phenomenon is named the *phase flip* in this paper. The oscillation frequency causing the phase flip is named the flipping frequency, denoted as $f_{flip}$.

$$F_{gain}\cos\theta = -F_{gain}\cos(\theta + \pi) \quad (7)$$

Similarly, the sine term $\sin(\pi f_{os}N/f_1)$ will keep crossing zeros as $f_{os}$ increases, and phase flips will occur repeatedly such that the DFT-measured oscillation phase is switching between in-phase and out-of-phase conditions compared to the true oscillation phase, depending on the sign of $F_{gain}$.

### D. DFT phasor with different window lengths

When applying DFT using a longer window, i.e., greater $N$, to capture the oscillation, the term $\sin(\pi f_{os}N/f_1)$ in (5) could cross zero more easily, i.e., experiencing the phase flip at a lower oscillation frequency $f_{os}$.

IEEE Std C37.118.1-2011 [14] specifies the performance criteria, i.e., the latency and the total vector error, and defines P-class (protection) and M-class (measurement) PMUs. In practice, usually a 1-cycle DFT is used by P-class PMUs to get fast response, while a 5-cycle DFT by M-class PMUs to meet steady-state accuracy and noise/harmonic rejection. Therefore, oscillations captured by M-class PMUs are more heavily attenuated and tend to encounter the phase flip more easily than P-class PMUs.

### E. Applicability and limitation of DFT-based Phasor

The applicability and limitation of DFT-based phasor in capturing oscillations are determined by its accuracy in extracting the underlying oscillation properties from waveform data, which is quantified by the measurement gain $F_{gain}$ in (5). Fig. 3 depicts how $F_{gain}$ changes with $f_{os}$ and $N$, illustrating the analysis in the previous subsections but with an overall view. A few observations can be made.

- All curve segments with $F_{gain}<0$ mean that the oscillation captured by the DFT-based phasor has an opposite oscillation phase compared to the ground truth.
- With a larger $N$, the $N$-cycle DFT-based phasor encounters the phase flip at a lower oscillation frequency. The lowest frequency causing the phase flip is summarized in Table I for $N$=1, 2, …, 6.

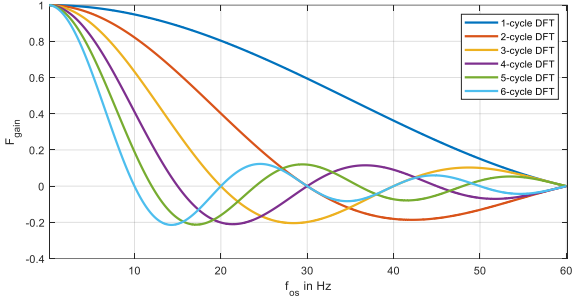

Fig. 3. Measurement gain $F_{gain}$ vs. $f_{os}$ and $N$

Table I. Lowest Flipping Frequency vs. DFT Window Length

| $N$ | 1 | 2 | 3 | 4 | 5 | 6 |
|---|---|---|---|---|---|---|
| $f_{flip}$ | 60 Hz | 30 Hz | 20 Hz | 15 Hz | 12 Hz | 10 Hz |

The phase flip is a phenomenon and limitation that is inherited from DFT. In practical applications, this limitation should be avoided by either carefully designing the DFT window length $N$ or replacing DFT with a non-DFT algorithm that does not have such a limitation. Next Section will introduce a multi-step estimation method that is free of the phase flip.

For existing PMUs that use DFT algorithms, it is important to realize their accuracy and applicability in capturing oscillations at different frequencies. Based on $F_{gain}$ in (5) and assuming, Table II provides a summary of the applicability, i.e., the accuracy of the DFT-measured oscillation amplitude with different DFT window lengths for oscillations at different frequencies. Table II shows that DFT-based phasors are fairly accurate for low-frequency oscillations. However, for high-frequency oscillations, a longer DFT window leads to increased attenuation of the oscillation amplitude and a higher likelihood of phase flip. Specifically, if one wants <10% error in estimated oscillation amplitude, then DFT window length should take a value for an oscillation at 7.4 Hz, then DFT window length cannot exceed two cycles, therefore, M-class PMUs typically won't work even regardless of the potential impact from LPF2.

Table II. Applicability of DFT-Based Phasor in Capturing Oscillations

| $f_{os}$ | $N$=1 | $N$=2 | $N$=3 | $N$=4 | $N$=5 | $N$=6 |
|---|---|---|---|---|---|---|
| 0.2Hz | 100% | 99.99% | 99.98% | 99.97% | 99.95% | 99.93% |
| 2.5Hz | 99.7% | 99% | 97% | 96% | 93% | 90% |
| 3Hz | 99.5% | 98% | 96% | 93% | 90% | 86% |
| 3.7Hz | 99.3% | 97% | 94% | 90% | 85% | 79% |
| 5Hz | 99% | 95% | 90% | 83% | 74% | 64% |
| 7.4Hz | 97% | 90% | 79% | 64% | 48% | 31% |
| 14Hz | 90% | 67% | 36% | 7% | - | - |
| 20Hz | 80% | 40% | 0% | - | - | - |

### F. Timestamping DFT-based phasors considering oscillations

For the waveform signal in (1), the amplitude-modulated oscillation in (2) is estimated by DFT to be (4) using time window [$\tau$, $\tau+N/f_1$]. To guarantee that the DFT-estimated oscillation phase in (4) is the same as the ground truth in (2), equation (8) must be satisfied, which requires that the timestamp of the DFT-based phasor must be the center of the DFT window in order not to introduce any error in the estimated oscillation phase. Such a requirement coincides with the recommendation by Standard [14].

$$2\pi f_{os}t + \varphi_2 = 2\pi f_{os}\tau + \frac{\pi f_{os}N}{f_1} + \varphi_2 \Rightarrow t = \tau + \frac{N}{2f_1} \qquad (8)$$

To illustrate the importance of the choice in the above for timestamping DFT phasors, Fig. 4 considers a signal with $A$=2, $m$=20%, $f_{os}$=5Hz, $\varphi_1$=$\varphi_2$=0 and two timestamping choices, one uses center of the DFT window represented by the green curve, and the other uses the ending time represented by the black curve. It is shown in Fig. 4 that assigning the timestamp at the end of the DFT window introduces a constant shift in the estimated oscillation phase, which is undesired.

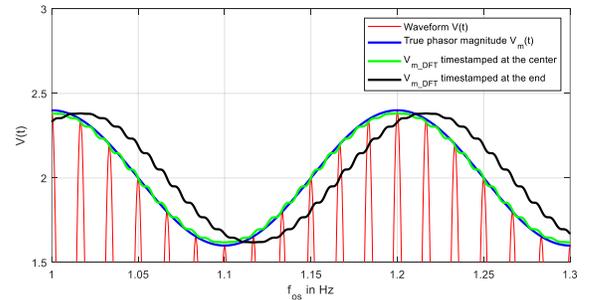

Fig. 4. One-cycle DFT phasor with different timestamping choices

## IV. A Multi-Step Parameter Estimation Method

To resolve the limitation of DFT algorithm shown in the previous Section, this Section proposes a novel multi-step parameter estimation method. The proposed method can be used as an alternative data processing for the existing PMU to produce phasors.

## A. Mathematical model of waveform signals

Consider a general waveform signal with modulated oscillations in both amplitude and angle as shown in (9). This signal model is adopted to account for the facts that (i) sub-synchronous component may have a different amplitude than that of the super-synchronous component, and (ii) oscillations can be simultaneously modulated in both amplitude and phase of the waveform at the fundamental frequency [12].

$$x(t) = a\cos[(2\pi f_1 t + \varphi_1 + h\cos(2\pi f_{os} + \varphi_{os}))] +$$
$$b_1\cos[2\pi(f_1 - f_{os})t + \varphi_{sub} + h\cos(2\pi f_{os} + \varphi_{os})] +$$
$$b_2\cos[2\pi(f_1 + f_{os})t + \varphi_{sup} + h\cos(2\pi f_{os} + \varphi_{os})] \quad (9)$$

All parameters in (9) are assumed to be unknown and need be estimated using PoW measurements, including $f_1$, $\varphi_1$, $a$, $f_{os}$, $\varphi_{os}$, $h$, $\varphi_{sub}$, $b_1$, $\varphi_{sup}$, and $b_2$.

## B. A multi-step parameter estimation method

Note that the least squares problem directly formulated using (9) is inherently nonlinear, and its solution may diverge if the initial guess of parameters is not sufficiently close to its true values. A multi-step method is proposed below to provide reliable estimates by leveraging an intermediate signal model shown in (10), which is a special case of (9) when $h=0$. After the substitutions in (12)-(13), the simplified signal model in (10) is equivalent to (11).

$$y(t) = a\cos(2\pi f_1 t + \varphi_1) + b_1\cos[2\pi(f_1 - f_{os})t + \varphi_{sub}] + b_2\cos[2\pi(f_1 + f_{os})t + \varphi_{sup}] \quad (10)$$

$$y(t) = A_1\cos(2\pi f_1 t) + B_1\sin(2\pi f_1 t) + A_2\cos[2\pi(f_1 - f_{os})t] + B_2\sin[2\pi(f_1 - f_{os})t] + A_3\cos[2\pi(f_1 + f_{os})t] + B_3\sin[2\pi(f_1 + f_{os})t] \quad (11)$$

$$a = \sqrt{A_1^2 + B_1^2}, b_1 = \sqrt{A_2^2 + B_2^2}, b_2 = \sqrt{A_3^2 + B_3^2} \quad (12)$$

$$\varphi_1 = \arctan\left(\frac{B_1}{A_1}\right), \varphi_{sub} = \arctan\left(\frac{B_2}{A_2}\right), \varphi_{sup} = \arctan\left(\frac{B_3}{A_3}\right) \quad (13)$$

If $f_1$ and $f_{os}$ in (11) are given, whose estimations will be shown later, then the signal model in (11) allows the formulation of a linear least square problem for estimating $A_1$, $B_1$, $A_2$, $B_2$, $A_3$ and $B_3$. Then, substituting them into (12)-(13) can determine $a$, $b_1$, $b_2$, $\varphi_1$, $\varphi_{sub}$ and $\varphi_{sup}$. The linear least square problem formulated using $L$ samples is shown in (15).

$$\begin{bmatrix} F_1(t_1) & F_2(t_1) & F_3(t_1) & F_4(t_1) & F_5(t_1) & F_6(t_1) \\ F_1(t_2) & F_2(t_2) & F_3(t_2) & F_4(t_2) & F_5(t_2) & F_6(t_2) \\ \vdots & \vdots & \vdots & \vdots & \vdots & \vdots \\ F_1(t_L) & F_2(t_L) & F_3(t_L) & F_4(t_L) & F_5(t_L) & F_6(t_L) \end{bmatrix} \begin{bmatrix} A_1 \\ B_1 \\ A_2 \\ B_2 \\ A_3 \\ B_3 \end{bmatrix} = \begin{bmatrix} y(t_1) \\ y(t_2) \\ \vdots \\ y(t_L) \end{bmatrix} \quad (15)$$

where $F_1(t)=\cos(2\pi f_1 t)$, $F_2(t)=\sin(2\pi f_1 t)$, $F_3(t)=\cos[2\pi(f_1+f_{os})t]$, $F_4(t)=\sin[2\pi(f_1+f_{os})t]$, $F_5(t)=\cos[2\pi(f_1-f_{os})t]$, $F_6(t)=\sin[2\pi(f_1-f_{os})t]$.

Key steps in the proposed parameter estimation method are summarized below.

**Step 1**. Estimate the fundamental frequency $f_1$ using the resampling-based DFT method in [16].

**Step 2**. Apply one-cycle DFT multiple times respectively to hopping windows of PoW data to obtain the estimate of phasor amplitude at multiple time instances, resulting in a phasor amplitude curve. The sliding windows should cover at least 1-2 periods of the oscillations. Guaranteeing this may require a few iterations with step 3.

**Step 3**. Apply Matrix Pencil method [18] [19] to the phasor amplitude curve from step 2 to obtain the oscillation frequency $f_{os}$ of the dominant oscillation mode.

**Step 4**. With $f_1$ from step 1 and $f_{os}$ from step 3, formulate the linear least square problem in (15) and solve for $A_1$, $B_1$, $A_2$, $B_2$, $A_3$ and $B_3$.

**Step 5**. Substitute estimated parameters from step 4 into (12)-(13) and obtain $a$, $b_1$, $b_2$, $\varphi_1$, $\varphi_{sub}$ and $\varphi_{sup}$.

**Step 6**. With parameters estimated from steps 1, 3 and 5, formulate a nonlinear least square problem using (9) with two unknown parameters, i.e., $h$ and $\varphi_{os}$. Then, solve it by a nonlinear solver such as the lsqcurvefit function in Matlab with the initial guess $h=\varphi_{os}=0$.

**Remarks**

- The above multi-step parameter estimation is proposed in this paper mainly to show the possibility that advanced signal processing techniques can overcome the limitations of DFT. The authors do not expect the proposed method to be the best for this estimation task. Further investigations and innovations are needed to develop accurate and computationally-efficient estimation algorithms for waveforms with high-frequency oscillations [20].

- It is also worth noting that if PoW measurements are available, the three-phase quantities can be transformed point to point into $\alpha\beta 0$ components, and subsequently into magnitude and phase [21]

- Assuming $h=0$ in step 4 simplifies the formulation to a linear least square (LLS) problem, as shown in (15). The LLS solution is unique and can be obtained with a single pseudo-inverse operation. Extensive testing shows that the six parameters estimated from (15) are consistently close to their ground-truth values, providing an excellent initial guess for the full nonlinear problem in step 6. The initial value of $h$ is set to zero, based on PSCAD simulations indicating that $h$ is typically small, which improves the likelihood of convergence. This approach has performed reliably across all tested signals in this paper.

## C. Error quantification

To quantify the error of an estimation method, the estimated parameters are used to reconstruct the waveform signal and then compare with the original waveform signal measurement. The following error metric is proposed and used throughout this paper, whose physical meaning is the average percentage error over a specified period. It is worth noting that the proposed error metric (17) simplifies to the total vector error [14] in the case of purely sinusoidal waveforms.

$$E_{LS} = \sum_{k=1}^{M}|y(t_k) - \hat{y}(t_k)|^2 \quad (16)$$

$$E_{PoW} = \frac{1}{a}\sqrt{\frac{\sum_{t_k=T_1}^{t_k=T_2}[\hat{y}(t_k) - y(t_k)]^2}{T_2 - T_1}} t_s \times 100\% \quad (17)$$

where $\hat{y}(t_k)$ and $y(t_k)$ represent the estimated and the measured waveform signals, respectively. $t_k$ represents the time associated with the $k$th sample, where $k$ denotes the time stamp index. $M$ represents the number of samples. $t_s$ is the sampling interval. $T_1$ and $T_2$ represent the start time and end time of the window used by the estimation algorithm, respectively. In the synthetic signal, the parameter $a$ is the ground truth of phasor amplitude as shown in (9), whereas in the real-world signal, $a$ is unknown and needs to be estimated.

Additionally, it is known that the least squares error (16) aims to minimize the difference between the estimated and modeled quantities when the least squares method is applied. To facilitate understanding of the error across different scales, the relationship between the least squares error and the proposed error metrics is presented as follows.

$$E_{\text{PoW}} = \frac{\sqrt{E_{LS}/M}}{a} \quad (18)$$

### D. Numerical example

To show the effectiveness of the proposed method on waveform signals that contain high-frequency oscillations, we test the signal in (9) with the following parameters: $f_1$=60Hz, $\varphi_1$=$\pi/3$, $a$=2, $f_{os}$=40Hz, $\varphi_{os}$=$\pi/6$, $h$=0.1, $\varphi_{sub}$=$3\pi/10$, $b_1$=0.25, $\varphi_{sup}$=$11\pi/30$, $b_2$=0.25. The sampling rate is 7.68 kHz.

Fig. 5 shows three waveform signals, where the red curve represents the original waveform in (9), the dashed blue curve represents the waveform reconstructed using one-cycle DFT-based phasor, and the black curve represents the waveform reconstructed using the proposed multi-step method. The error metric (17) of the reconstructed waveforms is 9% for one-cycle DFT method, while $1.32\times10^{-4}$% for the proposed method. Such a high accuracy of the proposed method can be seen in Fig. 5 by the fact that the reconstructed waveform basically overlaps with the original waveform. Similar observations can be made on the phasor amplitude curves, as shown in Fig. 6.

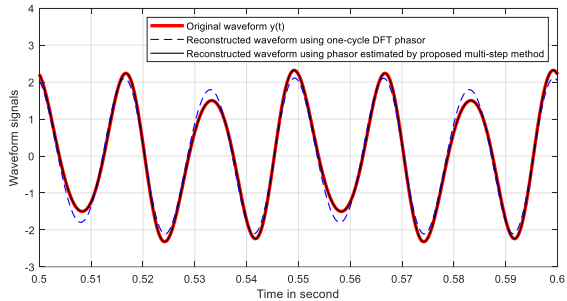

Fig. 5. Waveform comparison

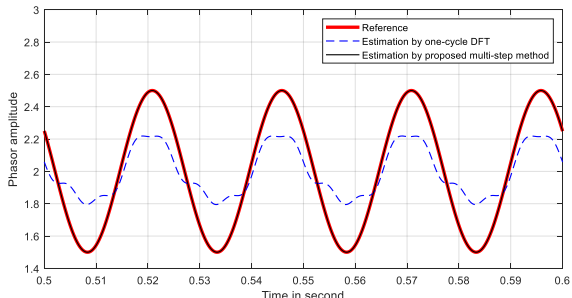

Fig. 6. Phasor comparison

## V. APPLICABILITY OF PHASOR CONCEPT

Previous sections show that the DFT-based phasor may become inaccurate for waveforms with high-frequency oscillations. Last section shows that using a specially designed parameter estimation method, the waveform with high-frequency oscillations could be accurately represented by the signal model in (9). This section aims to address the following question: is the signal in (9) always representable by a phasor?

The concept of phasor is a mathematical tool traditionally used to analyze linear circuits operating under sinusoidal steady-state conditions [22]. Power systems experiencing oscillations are apparently not under a sinusoidal steady-state condition. For simplicity, we consider the waveform signal in (9) with $h$=0, which is equivalent to the signal in (10). The following analyzes the signal in (10) under two different conditions: (i) $b_1$=$b_2$, and (ii) $b_1$≠$b_2$.

When $b_1$=$b_2$, the waveform signal in (10) becomes (19) using (20)-(21) if (22) holds true.

$$y(t) = [a + b\cos(2\pi f_{os}t + \varphi_2)]\cos(2\pi f_1 t + \varphi_1) \quad (19)$$

$$b = 2b_1 \quad (20)$$

$$\varphi_2 = \varphi_{sup} - \varphi_1 \quad (21)$$

$$\varphi_{sub} + \varphi_{sup} = 2\varphi_1 + 2k\pi, \ \exists k \in Z \quad (22)$$

In practice, once all parameters in signal model (10) are estimated and if $b_1$=$b_2$ and if (22) holds true, then the waveform signal in (10) is phasor representable as shown in (19) where parameters $b$ and $\varphi_2$ are defined in (20)-(21).

When $b_1$≠$b_2$, the waveform signal in (10) seems not phasor representable. Two numerical cases are presented below to demonstrate this. Consider a waveform signal (10) with the following parameters: $f_1$=60Hz, $\varphi_1$=$\pi/3$, $a$=2, $f_{os}$=2.8Hz, $\varphi_{sup}$=$11\pi/60$, and $\varphi_{sub}$=$-3\pi/20$. In case 1, $b_1$=$b_2$=0.2, while in case 2, $b_1$=0.2, $b_2$=0.1.

In each of cases 1 and 2, the waveform signal is generated using the signal model in (10). Then, the phasor-representable signal model in (19) is used to fit the generated curve, and the Gauss-Newton method is adopted to find the optimal parameters in (19).

To assess whether the waveform signal in (10) is phasor-representable, the error metric defined in (17) is applied. The estimation error of the phasor-representable signal in (19) is $9\times10^{-13}$% in Case 1, effectively reaching the limit of double-precision accuracy. In Case 2, a substantially larger error of 2.5% is observed. This error may become more significant for high-frequency oscillations caused by IBRs [23].

## VI. CASE STUDIES

In this section, voltage waveform signals are simulated in electromagnetic transient simulations and used to further verify the analyses and findings in this paper.

### A. DFT phasor evaluation on IEEE 9-bus system with IBRs

The PSCAD model of the IEEE 9-bus power system from PyPSCAD [24] is adopted, as shown in Fig. 7, which has two grid-forming (GFM) inverters, one grid-following (GFL) inverter, and no synchronous machine. After the system is initialized and stabilized at a steady state, three cases are simulated with different configurations to respectively excite 3-Hz, 8.4-Hz and 37.25-Hz oscillations. Phase-A voltage waveform at bus 2 is recorded and analyzed. The numerical integration step is 50 μs, and the plotting step is 500 μs.

- 3-Hz oscillation case: line 8-9 is opened.
- 8.4-Hz oscillation case: GFM-2's PLL $K_i$ gain is changed from 415 to 3540.
- 37.25-Hz oscillation case: GFM-2's PLL $K_p$ gain is changed from 52 to 1000.

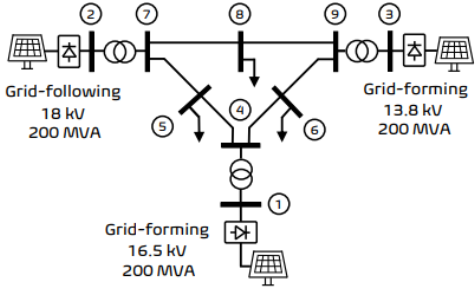

Fig. 7 One-line diagram of the modified IEEE 9-bus system (from reference [24], Fig. 3)

Bus 2's phase-A voltage waveforms in the three cases are shown in Fig. 8, where the data in the zoom-in subplots in blue are used to compute the DFT spectrum as shown in Fig. 9. Two observations on Fig. 9:

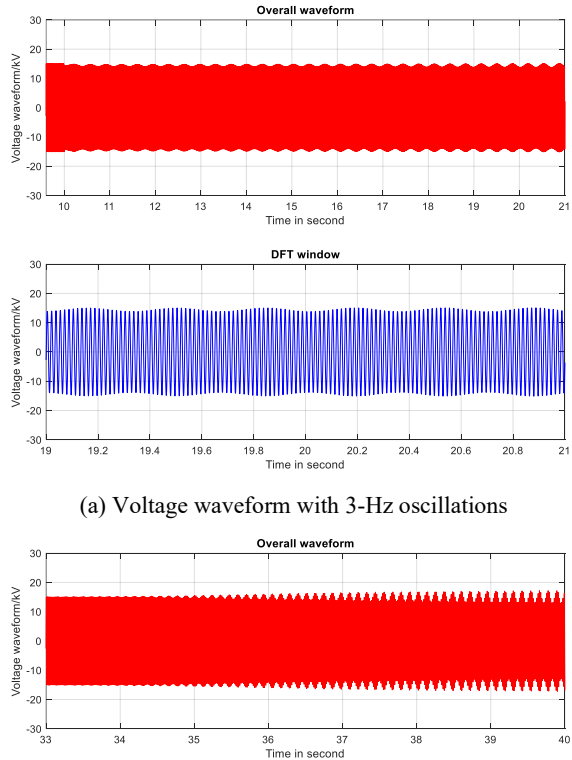

(a) Voltage waveform with 3-Hz oscillations

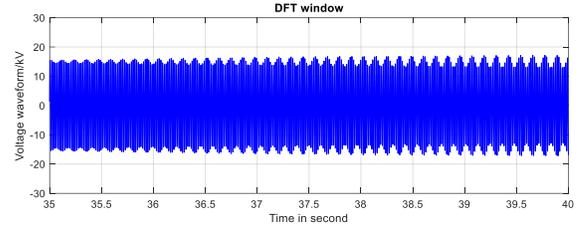

(b) Voltage waveform with 8.4-Hz oscillations

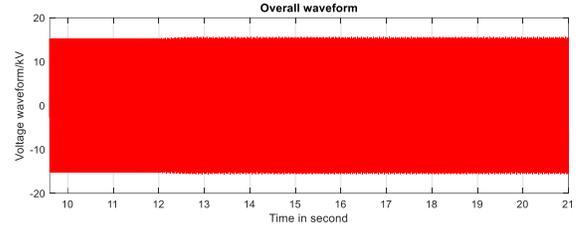

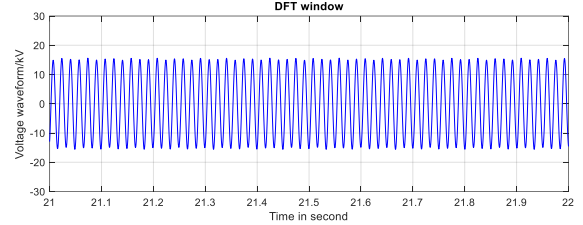

(c) Voltage waveform with 37.25-Hz oscillations

Fig. 8 Phase-A voltage waveforms at bus 2 respectively with 3-Hz, 8.4-Hz, and 37.25Hz oscillations.

- The frequency components are distributed approximately symmetrically around synchronous frequency, i.e., 60 Hz.
- For each sub- and super-synchronous frequency pair, their amplitudes are not equal. According to Section V-II, strictly speaking, all these waveform signals are not phasor-representable.

After applying a two-cycle DFT to compute the phasor amplitude curve and comparing with 30-fps PMU data with and without LPF, the results in Fig. 10 can be obtained. Regardless of potential attenuation caused by the two-cycle DFT algorithm, the blue, red and black phasor amplitude curves in Fig. 10 remain fairly accurate when characterizing 3-Hz and 8.4-Hz oscillations. However, for the 37.25-Hz oscillations, the 30-fps PMU data prior to LPF lack sufficient resolution, causing the oscillation to alias to 7.25 Hz, as shown by the red curve. In addition, the LPF, which is designed to attenuate high-frequency components and prevent aliasing, inadvertently distorts the red curve into the black curve, which corresponds to the PMU data typically measured in the field.

The PMU data with and without LPF, i.e., black and red curves in Fig. 10, are processed using the DFT algorithm to generate the DFT spectrum, as shown in Fig. 11. Similar observations can be made. Given that the PMU has a 30-fps reporting rate, the 37.25-Hz oscillation is aliased to 7.25 Hz. Such an aliasing phenomenon is well known and has been reported in a real-world oscillation event from Dominion [25]. Note that for the same oscillation event, oscillation frequency appears differently in the PMU and PoW data (see Appendix B). Therefore, when referring to the oscillation frequency, it is best practice to specify whether it is obtained from PMU data or PoW data.

Reconstructing the waveforms from the DFT-based PMU phasors for signals containing 3-Hz, 8.4-Hz, and 37.25-Hz oscillations, and computing their errors using (17) yield values of 10.4%, 8.1%, and 9.3%, respectively, whereas the corresponding estimation errors by the proposed multi-step method are 1.86%, 2.89%, and 3.14%.

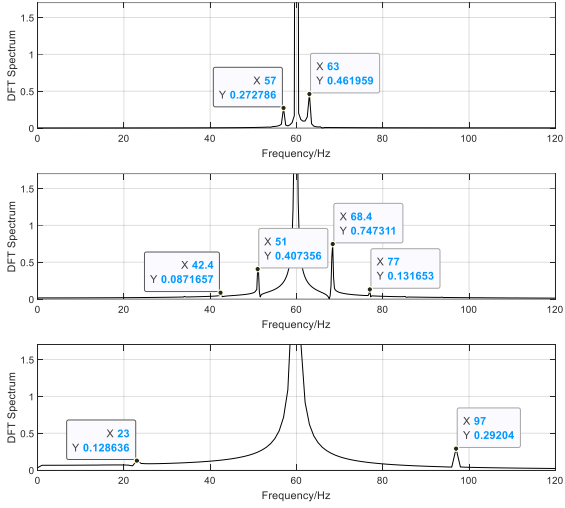

Fig. 9. DFT spectrum of waveform signals with 3-Hz (top), 8.4-Hz (middle) and 37.25-Hz (bottom) oscillations

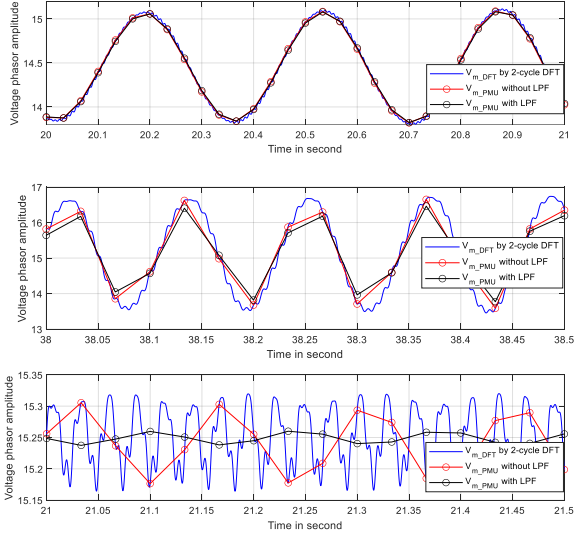

Fig. 10. Voltage phasor amplitude curves by two-cycle DFT on waveform signals with 3-Hz (top), 8.4-Hz (middle) and 37.25-Hz (bottom) oscillations

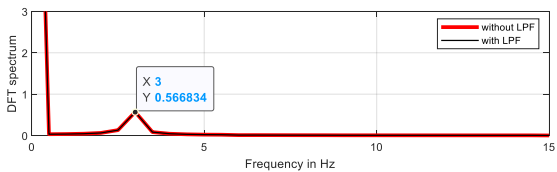

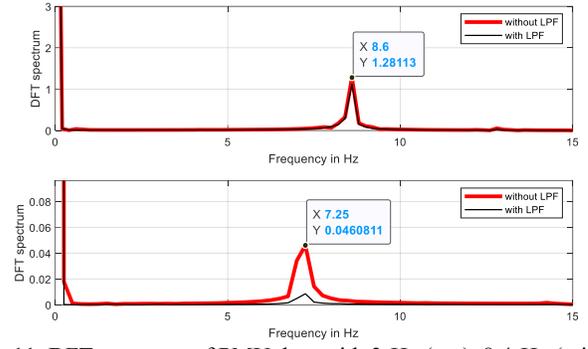

Fig. 11. DFT spectrum of PMU data with 3-Hz (top), 8.4-Hz (middle) and 37.25-Hz (bottom) oscillations

### B. Remarks on practical implications

Since DFT-based PMU phasors cannot accurately represent high-frequency oscillations superimposed on waveform signals, especially when the waveforms are not phasor-representable, one of the following two measures is required for PMU measurement-based oscillation analysis: (1) improve PMU algorithms, or (2) directly use PoW measurements.

To improve PMU algorithms, several enhancements are required. First, a more general signal model, such as the one in (9), is needed to represent waveforms that are not phasor-representable. Second, advanced estimation algorithms must be developed to optimally identify the parameters of this model. Instead of reporting phasor magnitude and angle as in conventional PMUs, the estimated parameters should then be transmitted over the communication network, which may require upgrades to accommodate the increased data volume. Finally, PMU-based oscillation analysis methods must be adapted or extended to process these estimated parameters. Collaborations among PMU manufacturers, PMU application software vendors, and working groups responsible for standard development will be essential to achieve this goal.

It should be noted that existing PMUs employing DFT-based algorithms and the wide-area measurement system built upon them remain accurate, see Table II, and effective for analyzing low-frequency oscillations. If improving PMU algorithms is not feasible, an alternative approach is to directly use PoW data that are measured by power quality meters. In this case, PMU-based oscillation analysis methods need to transition from phasor input to PoW data, which may not be a trivial process [26]

### VII. CONCLUSIONS

This paper presented a comprehensive analysis of the accuracy and applicability of DFT-based PMU phasors for oscillation measurements. Through theoretical derivations and numerical studies, we demonstrated that the DFT algorithm, anti-aliasing filters, and low reporting rates jointly distort high-frequency oscillatory components in point-on-wave (PoW) signals, leading to amplitude attenuation, phase flip, and aliasing. While these effects are negligible for low-frequency oscillations, they become significant as the oscillation frequency increases, limiting the reliability of conventional PMU measurements.

Simply increasing the reporting rate and adjusting LPF2 accordingly do not address the issue due to DFT algorithm. To overcome these limitations, a non-DFT multi-step estimation method combining one-cycle DFT, Matrix Pencil method, and least square error method was developed. The method achieves substantially improved accuracy in estimating waveform parameters and reconstructing both phasors, when possible, and PoW signals. The analysis further revealed that the traditional phasor concept itself may not be valid for waveforms containing asymmetric sub- and super-synchronous components, which are increasingly common in systems dominated by inverter-based resources.

These findings suggest that future PMU algorithm development should adopt generalized signal models and advanced parameter estimation techniques to better capture non-phasor-representable waveforms. When the improvement is impractical, the direct use of high-speed PoW measurements from power quality meters offers a viable alternative solution. In either case, waveform measurement systems and oscillation analysis methods must evolve accordingly.

## VIII. APPENDIX A

In this Appendix, the detailed steps are presented to derive the analytical form of DFT phasor for amplitude-modulated waveform signals in (1)-(2). First, the real and imaginary components of the phasor in (3) are obtained by applying the Fourier transformation, as follows:

$$X_r = \frac{2f_1}{N} \int_\tau^{\tau + \frac{N}{f_1}} A[1 + m\cos(2\pi \cdot f_{os} t + \varphi_2)] \cdot \cos(2\pi \cdot f_1 t + \varphi_1)\cos(2\pi \cdot f_1 t)dt \quad \text{(A-1)}$$

$$X_i = \frac{2f_1}{N} \int_\tau^{\tau + \frac{N}{f_1}} A[1 + m\cos(2\pi \cdot f_{os} t + \varphi_2)] \cdot \cos(2\pi \cdot f_1 t + \varphi_1)\sin(2\pi \cdot f_1 t)dt \quad \text{(A-2)}$$

The amplitude of the Fourier transform is determined by the 2-norm of its real and imaginary components:

$$X_{amp} = \sqrt{X_r^2 + X_i^2} = \sqrt{\left(A\cos\varphi_1 - \frac{2Af_1 m}{N} \cdot C\right)^2 + \left(A\sin\varphi_1 - \frac{2Af_1 m}{N} \cdot D\right)^2}$$

$$= A\sqrt{1 + \left(\frac{2f_1 m}{N}\right)^2 (C^2 + D^2) - \frac{4f_1 m}{N}(C\cos\varphi_1 + D\sin\varphi_1)} \quad \text{(A-3)}$$

where

$$C = \frac{\sin\left(\frac{\pi f_{os}N}{f_1}\right)}{f_{os}\pi(4f_1^2 - f_{os}^2)}\Big[f_{os}f_1 \sin\left(\varphi_2 + 2\pi f_{os}\tau + \frac{\pi f_{os}N}{f_1}\right)\sin(4\pi f_1\tau + \varphi_1) + f_{os}^2 \cos(2\pi f_1\tau + \varphi_1)\cos\left(\varphi_2 + 2\pi f_{os}\tau + \frac{\pi f_{os}N}{f_1}\right) - 2f_1^2 \cos\varphi_1 \cos\left(\varphi_2 + 2\pi f_{os}\tau + \frac{\pi f_{os}N}{f_1}\right)\Big] \quad \text{(A-4)}$$

$$D = \frac{\sin\left(\frac{\pi f_{os}N}{f_1}\right)}{f_{os}\pi(4f_1^2 - f_{os}^2)}\Big[f_{os}f_1 \sin\left(\varphi_2 + 2\pi f_{os}\tau + \frac{\pi f_{os}N}{f_1}\right)\cos(4\pi f_1\tau + \varphi_1) - f_{os}^2 \cos(2\pi f_1\tau + \varphi_1)\sin(2\pi f_1\tau)\cos\left(\varphi_2 + 2\pi f_{os}\tau + \frac{\pi f_{os}N}{f_1}\right) - 2f_1^2 \sin\varphi_1 \cos\left(\varphi_2 + 2\pi f_{os}\tau + \frac{\pi f_{os}N}{f_1}\right)\Big] \quad \text{(A-5)}$$

After neglecting the first term respectively in bracket in (A-4) and (A-5), which represent very fast ripples on the estimated phasor, the phasor amplitude can be approximated by the expression given in (4).

## IX. APPENDIX B

To illustrate how the same oscillation phenomenon is seen by PMU data and PoW data, this Appendix provides four plots displaying the same oscillation in (1) Fig. B1a for time-domain PoW, (2) Fig. B1b for frequency-domain PoW, (3) Fig. B1c for time-domain PMU, and (4) Fig. B1d for frequency-domain PMU. The PoW signal is created by using (1)-(2) with the following parameters: $A$=2, $m$=0.2, $f_1$=60 Hz, $f_{os}$=5 Hz, $\varphi_1$=π/4, and $\varphi_2$=π/5. PMU data is sampled at 30 fps.

The 5-Hz oscillation observed in PMU data is represented in PoW data by two components alongside the fundamental frequency: a 55-Hz sub-synchronous component, and a 65-Hz super-synchronous component. It is important to note that, in synchronous machine-dominated power systems, the frequency of low-frequency oscillations is typically referenced to the oscillation frequency values measured by PMUs, whereas sub-synchronous oscillation frequencies are generally referenced to those observed in PoW data.

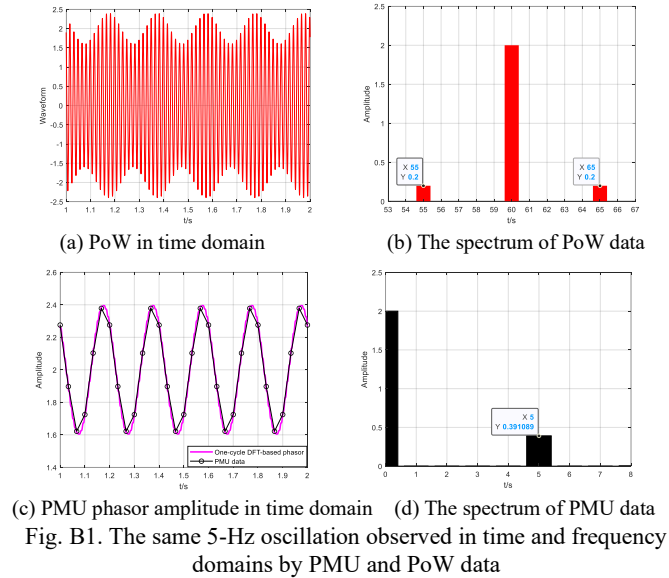

(a) PoW in time domain  (b) The spectrum of PoW data

(c) PMU phasor amplitude in time domain  (d) The spectrum of PMU data

Fig. B1. The same 5-Hz oscillation observed in time and frequency domains by PMU and PoW data

## X. ACKNOWLEDGEMENT

The authors gratefully acknowledge Dexter Newton of DOE's Solar Energy Technologies Office (SETO) for his guidance and support, and Dr. Wenpeng Yu of ISO New England for helpful discussions.